\documentclass{IOS-Book-Article}
\usepackage[utf8]{inputenc} 
\usepackage{hyperref}
\usepackage{hypernat}

\usepackage{mathptmx}

\usepackage{graphicx}
\usepackage{url}
\usepackage{amssymb}
\usepackage[dvipsnames]{xcolor}
\usepackage{enumerate}
\usepackage{algpseudocode, algorithm}

\usepackage{caption}
\usepackage{subcaption}
\usepackage{multirow}

\def\hb{\hbox to 10.7 cm{}}

\begin{document}

\pagestyle{headings}
\def\thepage{}

\begin{frontmatter}              % The preamble begins here.

%\pretitle{Pretitle}
\title{Who is this Explanation for? \\Human Intelligence and Knowledge Graphs for eXplainable AI}

\markboth{February 2020\hb}{February 2020\hb}
%\subtitle{Subtitle}

\author{\fnms{Irene} \snm{Celino}} \footnote{Pre-print version of the Book Chapter accepted in: Ilaria Tiddi, Freddy Lecue, Pascal Hitzler (eds.), Knowledge Graphs for eXplainable AI -- Foundations, Applications and Challenges. Studies on the Semantic Web, Volume 47, IOS Press, Amsterdam, 2020.}%
%\author[A]{\fnms{Irene} \snm{Celino}}%
%\thanks{Use this if needed.},
%\author[B]{\fnms{Second} \snm{Author}}
%and
%\author[B]{\fnms{Third} \snm{Author}}

\runningauthor{Irene Celino}
\address{Cefriel -- Politecnico di Milano\\Viale Sarca 226, 20126 Milano -- Italy}
%\address[A]{Cefriel -- Politecnico di Milano}
%\address[B]{Short Affiliation of Second Author and Third Author}

\begin{abstract}
%\textcolor{red}{Your abstract here.}
eXplainable AI focuses on  generating explanations for the output of an AI algorithm to a user, usually a decision-maker. Such user needs to interpret the AI system in order to decide whether to trust the machine outcome. When addressing this challenge, therefore, proper attention should be given to produce explanations that are \emph{interpretable} by the target community of users.

In this chapter, we claim for the need to better investigate what constitutes a \emph{human explanation}, i.e. a justification of the machine behaviour that is interpretable and actionable by the human decision makers. In particular, we focus on the contributions that \emph{Human Intelligence} can bring to eXplainable AI, especially in conjunction with the exploitation of Knowledge Graphs. 

Indeed, we call for a better interplay between Knowledge Representation and Reasoning, Social Sciences, Human Computation and Human-Machine Cooperation research -- as already explored in other AI branches -- in order to support the goal of eXplainable AI with the adoption of a \emph{Human-in-the-Loop} approach.
\end{abstract}

\begin{keyword}
Explainability \sep Human Intelligence \sep Human Computation \sep Human-in-the-Loop \sep Human-Machine Cooperation \sep Knowledge Graphs
\end{keyword}
\end{frontmatter}

\markboth{February 2020\hb}{February 2020\hb}
%\thispagestyle{empty}
%\pagestyle{empty}

%%%%%%%%%%%%%%%%%%%%%%%%%%%%%%%%%%%%%%%%%%%%%%%%%%%%%%%%%%%%%%%%%%%%%%%%%%%%%%%%%%%%%%%%%%%%%%%%
\section{Introduction}\label{sec:intro}

%Modeling (as in ML) needs humans (slide 23)
The recent renaissance of Machine Learning and Artificial Intelligence approaches brought a new wave of interest in such methods and technologies. Autonomous agents and automatic systems are now available and more affordable than before, but, if we relied only on popular news and communication, we would tend to think that they completely got rid of human intervention both in their setup and in their operation. Any practitioner, however, knows very well that human contributions are indispensable in order to set up, train, optimise and operate such systems.

Referring to the AI systems that more strongly rely on data and in particular to predictive Machine Learning, human knowledge is still required in all phases to answer relevant questions that are not necessarily targeted to the AI experts:
\begin{itemize}
	\item \emph{before creating a model}: 
		%\begin{itemize}
		%\item 
		during training set creation (``what data can I use to build a model?'')
	%\end{itemize}
	\item at \emph{model building time}: 
	%\begin{itemize}
		%\item 
		during model validation (``is my model correct?'', ``is my model good enough?'')  and 
		%\item 
		during model refinement (``what additional training data/features would improve my model performance?'')
	%\end{itemize}
	\item \emph{using the model in production}: 
	%\begin{itemize}
		%\item 
		to ensure algorithmic transparency (``should I trust the way my model gave such a prediction?'') and 
		%\item 
		to provide explainability (``why did my model give such an outcome/prediction?'')
	%\end{itemize}
\end{itemize}

\noindent In this chapter, we focus on the role that Human Intelligence and (human-generated) Knowledge Graphs play to answer the above questions. We also claim that, with special reference to explainability, humans are only partially considered in eXplainable AI research, while they should, because the required explanations should be useful for human comprehension.

The remainder of the chapter is structured as follows: related work is illustrated in Section \ref{sec:related}, and Section \ref{sec:explanation} clarifies what we mean by explanation and why humans are needed in their generation; opportunities for (human) eXplainable AI coming from the employment of Human Intelligence and Knowledge Graphs are outlined in Section \ref{sec:ciccia}, and Section \ref{sec:concl} presents some conclusions and traces some possible future work. %\textcolor{red}{TBC}

%%%%%%%%%%%%%%%%%%%%%%%%%%%%%%%%%%%%%%%%%%%%%%%%%%%%%%%%%%%%%%%%%%%%%%%%%%%%%%%%%%%%%%%%%%%%%%%%
\section{Related Work}\label{sec:related}

In the context of Artificial Intelligence and Machine Learning, several research trends investigate the role and interplay between humans and machines.

A new emerging process of scientific inquiry is shown in~\cite{shih2018beyond}: different people beyond scientists are now involved in such a process, because laymen participate both in the creation/collection of information (via user-generated content) and in the coding/labelling/validation phases (e.g. through Crowdsourcing or Citizen Science); the authors call for a new data analytics paradigm with user involvement, and demonstrate experimental results to show the effect of interface design on how users transform information.

Indeed, the power of the ``crowd'' is often leveraged to create large-scale training sets for Machine Learning, by adopting Crowdsourcing~\cite{howe2008crowdsourcing}, Human Computation~\cite{law2011human} and Citizen Science~\cite{irwin2002citizen} approaches.
Moreover, knowledge in human cognitive processes may assist the design and implementation of Machine Learning, as claimed in~\cite{zheng2018effective}; however, the current popularity of black-box models hinders an effective human intervention because those approaches negatively impact on trustworthiness, interpretability and the discovery of hidden rules.

User trust is indeed an important indicator because it correlates with system accuracy: humans are able to dynamically adjust their reliance based on a system perceived accuracy and they even show acceptance thresholds~\cite{yu2018trust}: this implies the need to correctly design an AI system to sustain the desired level of user trust.
The validation phase of Machine Learning algorithms also  benefits from the integration of user-centred evaluation: the authors of~\cite{cambo2018user} advocate adopting user-centred design (iterative) approaches for Machine Learning, in model optimisation, selection and validation.

Different families of methods explicitly aim to improve learned models based on human knowledge. Active Learning~\cite{settles2009active} is based on the idea that a Machine Learning algorithm can achieve greater accuracy with fewer labeled training instances if it is allowed to ``choose'' the data from which it learns, by asking queries to an ``oracle'' which usually is a human annotator. Transfer Learning~\cite{pan2009transfer} emerged to fulfill the need to build real world applications in which it is expensive or impossible to re-collect the  training data required and rebuild the models; in such cases knowledge transfer is attempted by adapting a model already trained on some domain (with the help of human annotators) to a different domain.

Human-Machine Cooperation is at the heart of Interactive Machine Learning~\cite{amershi2012interactive}, in which a human operator and a machine collaborate to achieve a task; while coupling algorithm-centred analysis with human-centred evaluation seems to yield better results than a fully automated or fully manual approach, research is still needed to explore to what extent this mix can provide benefits~\cite{boukhelifa2018evaluation}: participatory design with end-users could help incorporating human expertise in algorithms and models; visualisation techniques could facilitate user feedback; creativity, lateral thinking and exploration can also support, if suitable tools and objective and subjective metrics are developed.

In general, in order to improve and optimise the interaction between humans and machines, a perspective shift should be adopted, for example by walking away from a purely technical optimisation and embracing a designer mindset, like the one proposed in~\cite{noessel2017designing}, in which the author invites to stop seeing technologies as a collection of tools and gadgets and instead start seeing them as an evolutionary flow around human problems, whose parts ultimately integrate to create a new category of things named agentive technologies or ``AI that works for people''.

\section{What is an explanation for humans}\label{sec:explanation}

The rationale behind eXplainable AI research is that Artificial Intelligence systems should not only display an intelligent behaviour, but they also should be able to explain such behaviour. The naturally raising question is what an explanation is and how to generate it. In this section, we attempt at illustrating the characteristics of explanations and we justify the need for ``Human Intelligence'' and ``Human-in-the-Loop'' approaches also in relation to eXplainable AI.

%--------------------------------------------------------------
\subsection{A working definition of explanation}\label{sub:def-expl}

Let us consider the simple example of email categorisation between spam and non-spam. Here the task is binary classification (i.e., the output of a Machine Learning classifier is the labeling of each mail as spam or non-spam).

An explanation consists in a set of hints to understand the relationship between the characteristics of an individual (e.g. an email) and the model prediction on that individual (e.g. this email is spam). The explanation is used by a human decision-maker, who should decide whether to trust the system (e.g. accept or reject the prediction of the spam classifier)~\cite{ribeiro2016whytrust}.

User trust can happen at different levels: on the individual prediction, when the user requires an explanation about a specific instance (e.g. why \emph{this} mail is spam) or on an entire model, when the user needs to decide whether to trust the system altogether. In the latter case, an explanation could require selecting a representative sample of individuals (e.g. a set of spam/non-spam emails) and explaining each individual in the sample.

The main characteristics that an explanation should display (again according to~\cite{ribeiro2016whytrust}) are fidelity, model-independence and interpretability. \emph{Local fidelity} or local faithfulness means that a prediction should be valid in the vicinity of the individual; global fidelity would of course be desirable, but it could be challenging for complex models. The explanation should also be \emph{model-agnostic}, in that it should be independent on the specific type of AI model. Finally, \emph{interpretability} is the qualitative understanding of the relationship between the input variables and the response (e.g. the relation between the words contained in an email and the email categorisation as spam/non-spam).

Interpretability is the key aspect for an explanation to be accepted by a user; in our example of an email classifier, an interpretable explanation could rely on a list of words (e.g. the system thinks this email is spam because it contains the following words) rather that be based on opaque clues (e.g. word embeddings) which are not easily understandable by a human. The level of interpretability of an explanation of course depends on the audience, because humans use their previous knowledge about the application domain to interpret an explanation and accept/reject a prediction based on their understanding.

%\cite{ribeiro2016whytrust} (slide 25)
%
%\begin{itemize}
	%\item Explanation = set of hints to understand the relationship between the characteristics of an individual (e.g. an email) and the model prediction on that individual (e.g. this email is spam)
	%\item Different levels of prediction trust
	%\begin{itemize}
		%\item On individual prediction: it requires explanation about the individual (e.g. why this mail is spam)
		%\item On entire model: it requires (1) selecting a representative sample of individuals (e.g. a set of spam/non-spam emails) + (2) explaining each individual in the sample
	%\end{itemize}
	%\item Characteristics of explanations
	%\begin{itemize}
		%\item Local fidelity: valid in the vicinity of the individual (but non necessarily globally)
		%\item Model agnostic: independent on the specific black box model
		%\item Interpretability: quantitative understanding of the explanation (e.g. words, not word embeddings); this depends on the audience, because humans use their previous knowledge to interpret an explanation
	%\end{itemize}
%\end{itemize}

%-----------------------------------------------------------
\subsection{Explanation from a human point of view}\label{sub:hum-expl}

The latest point on interpretability clarifies that proper attention should be given to the different kinds of explanations that could be generated, in particular by distinguishing ``machine'' explanations from ``human'' explanations.

Indeed, most XAI research has been focusing on generating \emph{machine explanations}, i.e. justifications of what/how the machine ``thinks''. In other words, machine explanations try to explain the scientific theory behind a model, to allow for phenomena comprehension. In case of interpretable models (like linear regression or decision trees), the machine explanation consists in making explicit the mathematical/logical relation between inputs and outputs (e.g. tree model of decisions). In case of black-box models, especially for deep learning and other complex approaches, the machine explanation may be based on ``compressed'' models or other approximation techniques that still use an explicit representation of the relation between inputs and outputs.

Instead, \emph{human explanations} focus on what a human user wants to know in order to interpret a model and make subsequent decisions. The user may be uninterested in the internal functioning of an algorithm, as she may even be unable to understand a potentially complex mathematical formulation of the function that transforms the input parameters in the output prediction. On the contrary, the user is interested in getting useful clues on why a specific output is given, in order to evaluate if such output is ``reliable'' from a human understanding point of view. 

As a consequence, in order to be useful, a human explanation needs to display some specific characteristics~\cite{mittelstadt2019explaining}. An explanation should be \emph{selective}: it should not provide all possible reasons, but convey only the ``relevant'' causes;  indeed, people usually do not expect an explanation to consist of a complete cause of an event, also to let the explanation itself being reduced to a cognitively manageable size; moreover, an explanation should not contain useless information, like presuppositions or beliefs that the user already holds. Humans psychologically prefer \emph{contrastive} explanations, in that they are used to reason according to counter-factual causality (i.e. people do not ask why an event A happened, but rather why an event A happened instead of some other event B), especially in case of an anomaly or an abnormal event. Another characteristic of human explanations is that they are usually \emph{social}, involving the interaction between (multiple) explainers and explainees; also with respect to eXplainable AI, explanations should be seen as an interactive process, including interaction and dialogue with a mix of human and machine participants.

From all the above considerations, it is apparent that eXplainable AI research should go well beyond automatic methods to generate explanations; it is of utmost importance to keep the \emph{Human-in-the-Loop}. There are at least two main reasons to advocate for the active involvement of people in eXplainable AI~\cite{miller2017beware}: on the one hand, if explanation formulation is delegated to ``computer scientists'', the risk is that such explanations are too close to the model and too far from human understanding, especially that of domain/business users who need to interpret such information; on the other hand, there is a large body of knowledge about explanations from the social sciences (philosophy, psychology, cognitive science), which could bring tangible benefits to eXplainable AI research in terms of getting to a ``good'' explanation from a human point of view~\cite{miller2019socialsciences}.

%\cite{mittelstadt2019explaining,miller2017beware} (slide 26)

%\begin{itemize}
	%\item Two meaning of explanation
	%\begin{itemize}
		%\item ``Machine'' explanation = what the machine thinks (scientific theory, phenomena comprehension)
		%\item ``Human'' explanation = what the human wants to know to interpret a model
	%\end{itemize}
	%\item Characteristics of explanations from the human point of view
	%\begin{itemize}
		%\item Selective explanations (not all possible reasons, but only ``relevant'' causes, not including pre-existing beliefs/assumptions)
		%\item Contrastive explanations (counterfactual causality, ``why P and not Q?'')
		%\item Social explanations (dialogue/conversation, interaction, iteration)
	%\end{itemize}
	%\item Why human-in-the-loop is needed for Explainable AI
	%\begin{itemize}
		%\item Don't let computer scientists decide how to formulate explanations, because otherwise explanations are too close to the model and too far from human understanding
		%\item There is a large body of knowledge about explanations from social sciences
	%\end{itemize}
%\end{itemize}

%%%%%%%%%%%%%%%%%%%%%%%%%%%%%%%%%%%%%%%%%%%%%%%%%%%%%%%%%%%%%%%%%%%%%%%%%%%%%%%%%%%%%%%%%%%%%%%%
\section{Human Intelligence and Knowledge Graphs to support eXplainable AI}\label{sec:ciccia} %why/how can humans and kg help xai

The Semantic Web has always relied on humans, since most of its tasks are knowledge-intensive and context-specific and, as such, they require user engagement for their solution (e.g., conceptual modelling, multi-language resource labelling, content annotation with ontologies, concept/entity similarity recognition). With the rise of Knowledge Graphs and their popularity, new opportunities have emerged to exploit them for AI in general and specifically for eXplainable AI~\cite{lecue2019kg4xai}.

\begin{figure}[htb]
	\centering
		\includegraphics[width=0.95\textwidth]{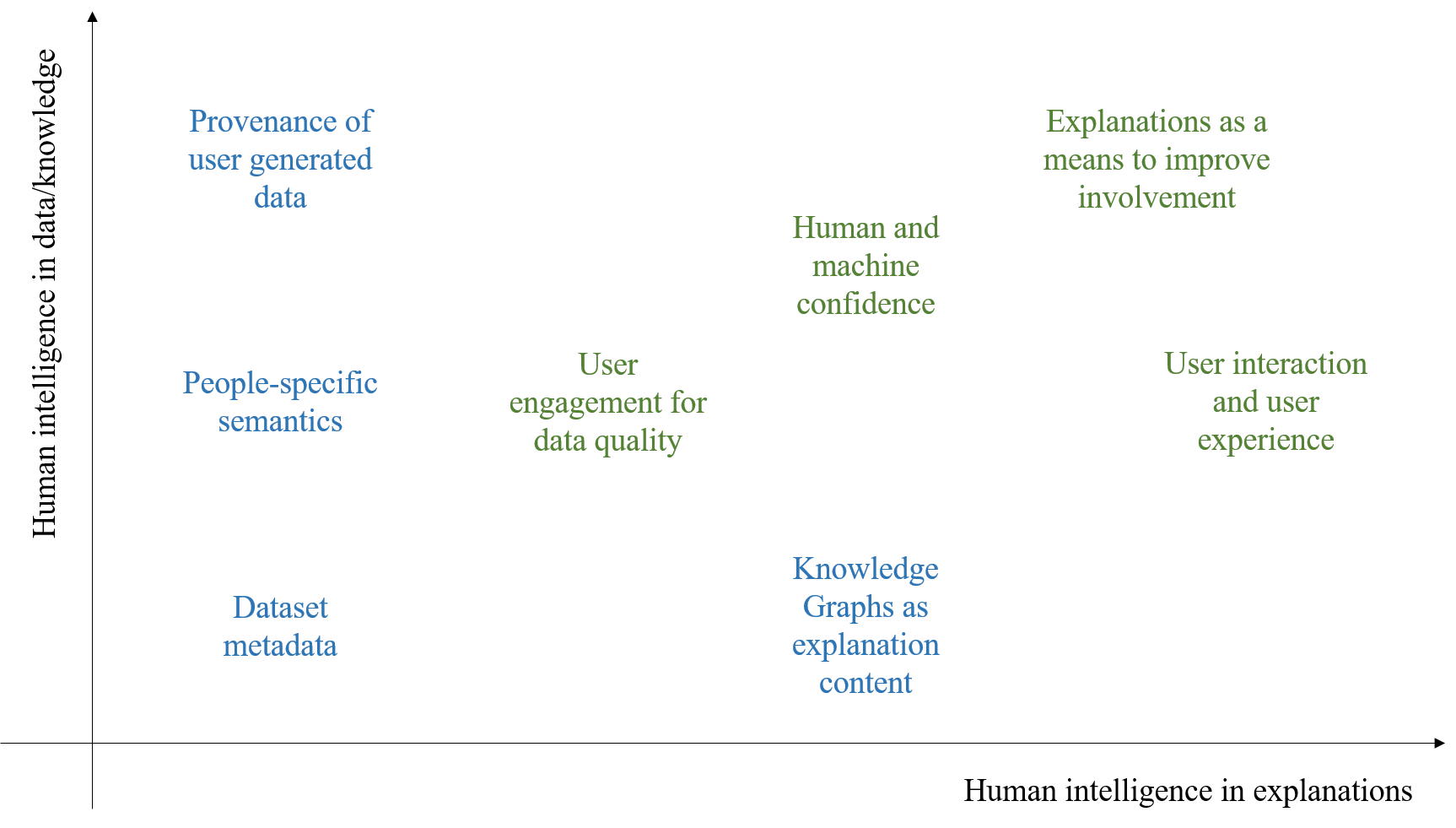}
	\caption{Graphical representation of Human Intelligence approaches}
	\label{fig:approaches}
\end{figure}

Without the claim of being exhaustive, in the following we illustrate a set of approaches that can bring Human Intelligence and Knowledge Graphs to the benefit of eXplainable AI, with specific reference to Machine Learning. We distinguish between two main types of opportunities, those related to the exploitation of (human-generated) Knowledge Graphs and those that capitalise on the direct involvement of people; we depict them in Figure~\ref{fig:approaches} along two axes, representing whether Human Intelligence is employed in data/knowledge representation or for explanations. %in relation to data preparation/manipulation and with respect to model explanation.

%\textcolor{red}{TODO: fare tante piccole sezioni e in ciascuna (1) definire a cosa serve per XAI e (2) spiegare meglio cosa sto raccontando/citando}

%-%-%-%-%-%-%-%-%-%-%-%-%-%-%-%-%-%-%-%-%-%-%-%-%-%-%-%-%-%-%-%-%-%-%
\subsection{(Human) Knowledge Graphs for XAI}
The first set of approaches exploits Knowledge Graphs to support explanation generation. We specifically focus on the role of human-generated information to directly and indirectly support XAI.

%-----------------------------------------------------------
\subsubsection{Dataset metadata}
Structured data represents an invaluable input for any Machine Learning approach. Consequently, linked data and Knowledge Graphs represent as such a rich and priceless contribution. An important role can even be played by simple metadata: descriptive metadata about datasets, in the form of DCAT~\cite{dcat-v2} and related vocabularies, can be exploited to improve data sourcing. The information about where some data comes from can also be re-used for explanations: users can better judge the reliability or the meaningfulness of a machine output if they are given also the detail about the original sources.

For example, the opportunities to facilitate dataset reuse in the development of chatbots are illustrated by the BotDCAT-AP vocabulary~\cite{cappello2017botdcat}, an extension of the Data Catalogue (DCAT) Application Profile. BotDCAT-AP enables the description of intents (i.e., the actions users want to accomplish by interacting with a chatbot) and entities (i.e., individual information units associated to an intent) supported by a dataset and the method to access it; as such, it enables and fosters reusability of datasets (including Knowledge Graphs) across chatbot systems. It could also be exploited further to support the generation of explanations for the chatbot ``replies'' in terms of recognised intents/entities and used datasets.

%-----------------------------------------------------------
\subsubsection{People-specific semantics}
Different users may have different interests or skills and, as a consequence, they may need different explanations. User-generated data often implicitly contains hints on \emph{what people care about}; this can be an opportunity to exploit when providing explanations on systems trained on such data.

For example, spatial data analytics of OpenStreetMap manual tagging showed to be beneficial for geo-ontology engineering, by surfacing latent semantic differences in concepts by different communities~\cite{recalegari2016supporting}: the same ``concept'' of spatial object (e.g., a pub) may have slightly diverging meanings in different places (e.g., a place to dine in UK, a bar to have a drink in Italy). This implicit semantics, when extracted and made explicit,  can be exploited also for explanation generation, because it can contribute to convey the right ``semantics'' to the right community.

%-----------------------------------------------------------
\subsubsection{Provenance of user-generated data}
Providing better data for training in turn leads to better models, as well as to more interpretable explanations. When data is user-generated, quality assurance is an important step, for example to aggregate inputs from multiple contributors (cf. ``truth inference'' in Crowdsourcing~\cite{zheng2017truth}). Provenance metadata about human contributions often contain important clues that can be exploited both for quality improvement and for generating explanations.

%When data is user-generated, provenance information can be exploited to improve data quality, as
For example, in the case of the Human Computation-powered volunteered geographic information (VGI) illustrated in~\cite{celino2013human}, the involvement of a crowd of volunteers, potentially untrained or non-experts, implies that VGI can be of varying quality; tracing VGI provenance enables the recording of the collection activity: the information about who gathered what, where and when is then employed to compute and judge the VGI quality. The same provenance information can be offered to users of systems trained on such user-generated data, to explain where some prediction comes from.

%-----------------------------------------------------------
\subsubsection{Knowledge graphs as explanation content}
Structured knowledge and Knowledge Graphs can be used as basis for explanations, because they may already contain the rationale behind the relationship between inputs and outputs of a system. Whenever a predictive system is based on a knowledge base, the relevant part of it that motivates a system output can be directly used as explanation.

%, in terms of resources and properties connecting them,
For example, graph traversal information is used to explain the suggestions of a knowledge-based recommender system in~\cite{dellaglio2010anatomy}: the logical path connecting a user (e.g., John loves hard rock music) and a recommended item (e.g., X is a Web-radio broadcasting rock music) provides a digestible account of the reasons behind the recommendation (e.g., John is recommended to listen to X, because John \emph{loves} hard rock music, hard rock \emph{is a kind of} rock music, X \emph{broadcasts} rock music). The chain of relevant connected resources/properties (i.e., the set of triples composing a path between the user and the recommended item) already constitutes a human explanation for the recommendation.

%-%-%-%-%-%-%-%-%-%-%-%-%-%-%-%-%-%-%-%-%-%-%-%-%-%-%-%-%-%-%-%-%-%-%
\subsection{Human Intelligence for XAI}
The second set of approaches directly focuses on the active involvement of people to the benefit of XAI.

%-----------------------------------------------------------
\subsubsection{User engagement for data quality}
As claimed in Section \ref{sec:explanation}, in order to provide human explanations, we should turn to social sciences, which may help in the involvement and engagement of people also during the phase of explanation generation for AI systems. Engaging humans is a challenge by itself, therefore eXplainable AI could reuse the research results in relation to designing and exploiting behaviours, personal motivations and incentive mechanisms. 

For example, the evaluation and improvement of data quality can be achieved through an analysis of contributions: %the unconscious contributions of users as well as by improving user experience It is undoubted that people can display varied levels of attention when performing a task, due to attitude or contextual conditions; therefore, 
user behaviour influences data quality and should be taken into account,  to evaluate the reliability of user-generated information, to better design data collection systems and to generate explanations. As demonstrated in~\cite{recalegari2018interplay}, the presence of tangible rewards, leveraging extrinsic motivation, affects quantity and quality of collected data; moreover, an analysis of accuracy and participation of contributors highlights different engagement profiles, which should be taken into account when aggregating user-generated data and should be capitalised for explainability. 

%-----------------------------------------------------------
\subsubsection{User interaction and user experience}
Lessons learned and best practices from user experience design can also inform human-powered explanation generation, because they can help in designing suitable tools and data value chains that involve and engage people to bring benefit to AI in general and eXplainable AI specifically.

Indeed, a carefully designed user interaction with digital tools proves to be key in raising attention and improving data quality, as shown in~\cite{celino2020submitting} with respect to survey data collection: an improvement on user experience, making questionnaire compilation more enjoyable, leads also to higher-quality information, because it reduces the satisficing effect and increases response quality. Therefore, involving users for the generation or validation of explanations, for example by adopting a social and interactive pattern guided by a design thinking approach, can maximise user attention and ease user experience, thus making sure that the result is a \emph{good} explanation from a human point of view.

%-----------------------------------------------------------
\subsubsection{Human and machine confidence}
Predictive Machine Learning modelling aims at building a trustworthy system able to provide prediction on unknown cases; to evaluate model confidence, different metrics are usually employed to give quantitative estimates of a prediction reliability. Reporting confidence metrics to support prediction explanation is a means to increase user trust, but again those quantitative hints should be interpretable from a human point of view.

Human intervention can be also employed to support a model evaluation and, consequently, a model explanation through confidence metrics. Indeed, it can happen that what is ``difficult'' to predict for an algorithm (i.e. predictions with low confidence metrics) is also difficult for humans to judge; the case of questionable image classification is illustrated in~\cite{recalegari2018human}, where the correspondence exists between low-confidence machine classifications and user disagreement. The correlation between human and machine predictions and their respective confidence/reliability can be exploited to understand the reasons behind a model and can therefore improve both the modelling phase (by incorporating additional human knowledge in training) and the generation of explanations (which can be closer to human understanding).

%-----------------------------------------------------------
\subsubsection{Explanation as a means to improve user involvement}
The most challenging aspect of Human-Machine Cooperation is the effective involvement of people in the various phases of  modelling. While users are already employed in data collection and model validation, further opportunities lie in a more interwoven interaction between human steps and automatic steps. Therefore, explanations are not only an objective as such, but they can be an instrument to further involve and motivate human participants in the AI system life-cycle.

For example, in order to identify and reduce bias in knowledge representation and modelling, the involved users should not only be exposed to potential biased information, but should also be given an explanation for such an identified bias, to understand the reasons behind a questionable piece of information or prediction. A Human-in-the-Loop approach to identify and resolve implicit bias in Knowledge Graphs is illustrated in~\cite{summerschool}: users are involved not only to accept/reject an identified bias, but they are also engaged as decision-makers to evaluate if further actions should be taken to solve such bias.

%-----------------------------------------------------------
%\subsection{}

%-----------------------------------------------------------
%\subsection{Opportunities related to Data for Explanation}\label{sub:data}
%\begin{itemize}
	%\item metadata to improve data sourcing (reusable for explanations?) \cite{cappello2017botdcat}: descrivere i dataset per facilitarne il riutilizzo??
	%\item provenance to evaluate and improve data quality \cite{celino2013human}: ????
	%\item ugc to understand what people care about \cite{recalegari2016supporting}: crowdsourced data give information in an implicit way??
	%\item user behaviour to evaluate and improve data quality \cite{recalegari2018interplay}: comportamento degli utenti influenza i risultati
	%\item user engagement to improve data quality \cite{celino2020submitting}: engagement degli utenti migliora i risultati?
%\end{itemize}

%-----------------------------------------------------------
%\subsection{Opportunities related to Model Explanation}
%\begin{itemize}
	%\item ugc to evaluate model confidence \cite{recalegari2018human}: quello che e' difficile per l'uomo e' difficile anche per la macchina (motivazione)
	%\item kg as explanation \cite{dellaglio2010anatomy}: path in the kg as en explanation for recommendations
	%\item provenance as explanation \cite{summerschool}: bias degli input utente e loro possibile tracciatura/provenance
%\end{itemize}

%%%%%%%%%%%%%%%%%%%%%%%%%%%%%%%%%%%%%%%%%%%%%%%%%%%%%%%%%%%%%%%%%%%%%%%%%%%%%%%%%%%%%%%%%%%%%%%%
\section{Conclusions}\label{sec:concl}

eXplainable AI aims at generating explanations to justify the output of an algorithm to a user, usually a decision-maker. Those explanations need to be interpretable by the intended target users and, therefore, cannot be restricted to the ``scientific modelling'' (i.e., the explanation of the scientific/mathematical law or theory behind an artificial model), but should be focused on addressing the needs of the decision makers, which exploit such explanations and decide whether to trust an AI system. 

Therefore, a better understanding of \emph{Human Intelligence} is needed to make sure that the generated explanations are ``good enough'' to be used in practice: a certain help can come from social sciences, but even within the ICT community, we identify several opportunities. On the one hand, Knowledge Representation and Reasoning (KRR) research has always been addressing the open issue of human knowledge formalisation; in this context, therefore, eXplainable AI can leverage all the experience related to the involvement of human annotators and crowdsourced knowledge bases and Knowledge Graphs (e.g. DBpedia~\cite{dbpedia} and Wikidata~\cite{wikidata}): indeed, the same Human Intelligence that supports KRR tasks can be similarly exploited for eXplainable AI.

On the other hand, Human Computer Interaction (HCI) research has been focusing on improving and optimizing user experience with digital tools; in this context, eXplainable AI can leverage the approaches and methods to support the ``interaction'' between a human user and a digital explanation, improving interpretability and promote trust. AI system should be designed to allow and facilitate the exchange with the relevant user communities: while people are already heavily involved in data collection, their engagement in other steps of the AI life-cycle is still to be fully explored, especially with respect to explainability.

The big challenge is to define flexible and complex human-computer cooperative systems, able to guide in the preparation, building and production of data processing pipelines involving Artificial Intelligence technologies. Human Intelligence and Knowledge Graphs should become first-order citizens of such data value chains, not only to improve the performance of such artificial systems, but also -- and foremost -- to assure that AI outcomes are relevant and usable by human decision makers.

\section*{Acknowledgments}
The presented research  was partially supported by the  ACTION project  (grant  agreement  number  824603),  co-funded  by  the European  Commission  under  the  Horizon  2020  Framework Programme. We would like to thank Gloria Re Calegari and Ilaria Tiddi for their feedback and revision on this chapter. \emph{}

\bibliographystyle{ios1} 
\bibliography{biblio}
	
%\begin{thebibliography}{99}
%
%\bibitem{r4}
%P.A. Deus, Ater hoc et filius et mater praestet nobis,
%\textit{Paterhoc} \textbf{66} (1993), 856--890.
%
%\end{thebibliography}
\end{document}